\documentclass[%
 reprint,
nofootinbib,
 amsmath,amssymb,
 aps,
]{revtex4-2}

\usepackage{graphicx}
\usepackage{dcolumn}
\usepackage{bm}
\usepackage{hyperref}

\begin{document}

\preprint{APS/123-QED}

\title{What Promotes Smectic Order: Applying Mean Field Theory to the Ends}

\author{David A. King}
\email{daviking@sas.upenn.edu}
\affiliation{Department of Physics and Astronomy, University of Pennsylvania, 209 South 33rd St., Philadelphia, PA, 19104.
}
\author{Randall D. Kamien}%
\affiliation{Department of Physics and Astronomy, University of Pennsylvania, 209 South 33rd St., Philadelphia, PA, 19104.
}
\date{\today}

\begin{abstract}
	Not every particle that forms a nematic liquid crystal makes a smectic. The particle tip is critical for this behaviour. Ellipsoids do not make a smectic, but sphero-cylinders do. Similarly, only those N-CB alkylcyanobiphenyls with sufficiently long ($N\geq 8$ carbons) alkane tails form smectics. We understand the role of the particle tip in the smectic transition by means of a simple two-dimensional model. We model sphero-cylinders by ``boubas'' with rounded tips, and ellipsoids by ``kikis'' with pointed tips. The N-CB molecules are modelled by a small body with a polymer tail. We find that rounded tips and longer polymer tails lead to a smectic at lower densities by making the space between layers less accessible, destabilizing the nematic. 
\end{abstract}

\maketitle


	\section{Introduction and Formulation}
	
	Onsager recognized that the geometry of particles affects the structure of their ordered phases \cite{Onsager1949TheParticles}.  The most remarkable thing about his insight is that the nematic phase is {\sl unremarkable}: any fluid of sufficiently anisotropic particles will form a nematic liquid crystal, where the particles are homogeneously distributed but have a preferential orientation.  However, not all such particles form a smectic-A phase, a phase with the same orientational order but with a periodic density modulation in the direction of alignment. This was noticed by Frenkel \cite{Frenkel1991, Frenkel1984}, who considered a system of parallel ellipsoids. Smectics have strong orientational order, so the particles may be assumed parallel without loss of generality. He argued that this system had no smectic phase because it could be mapped to a system of hard spheres in a way that preserves the thermodynamic properties by simply rescaling the lengths and momenta parallel to the ellipsoids. Hard spheres are only observed to exist in fluid or crystalline phases, so the ellipsoids can have no smectic phase. 
	
	This argument is extremely elegant, but leaves some open questions; what if the particle shape is only \textit{approximately} an ellipsoid so that the rescaling does not produce spheres? Are ellipsoids the only elongated particles that miss the smectic phase due to this symmetry? Sphero-cylinders have been observed in simulations to make smectics \cite{Stroobants1986}; what do they have that those particles without smectic phases do not? Some of these questions can be tackled using density functional theory and similar methods \cite{Lipkin1983ASmectics,Evans2007LiquidGeometry,Wittmann2014,Wittmann2017, Hosino1979, Mulder1987Density-functionalFluid, Taylor1989TheorySpherocylinders}, but this often results in complicated analyses and it is difficult to gain insight into the differences between different particle shapes. 
	
\begin{figure}\includegraphics[width=8cm]{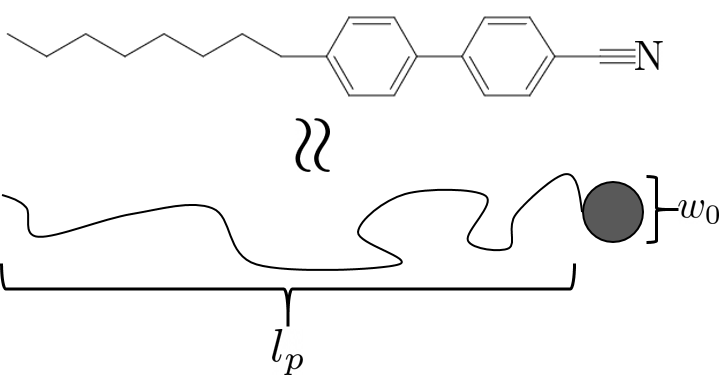}
		\caption{\label{fig:NCB} The chemical structure of the N-CB molecules ( specifically 8-CB) is given above our crude model for it. We think of these molecules as having a small body of size $w_0$ and a polymer tail of length $l_p$.}
	\end{figure}

	It is useful to look for another instance of two molecules with similar structures where one \textit{has} a smectic phase but the other \textit{does not}, so that similarities to the case of ellipsoids and sphero-cylinders can be sought. Such an example exists and is well known to experimentalists; $N$-CB type alkylcyanobiphenyls \cite{Gray1976, Cacelli2007}. The precise structure of these molecules is shown in Fig.(\ref{fig:NCB}), but it is most useful to think about them as a small ``body'' to which a ``tail'' made of $N$ links is attached. When $N=8$, the molecule is a typical thermotropic liquid crystal former, and has both a nematic and a smectic-A phase. With $N=5$, however, the smectic is absent (indeed, for $N<8$ there is no smectic, though most experiments focus on 5-CB). The common difference between the particles in the N-CB example and Frenkel's case is the structure at their ends; their ``tips''. This points to the key question we would like to answer: why are the particle tips important for the formation of a smectic phase?  We will argue here that the nematic phase is suppressed by rounded tips allowing the smectic phase to intervene.  This effect is similar to the situation found in \cite{deplete} where the introduction of small platelets suppressed the uniaxial nematic phase allowing for the onset of the biaxial nematic.  In short, when the mesogen tips are pointed, a test mesogen can more easily be inserted {\sl between} existing smectic layers compared to a round-tipped mesogen.   As a result pointy mesogens more easily fill in the space between smectic layers resulting in the nematic phase.  
	
	We tackle this problem by means of a toy model, which captures the essential physics but is simple enough to be understood fully. For this model to be satisfactory and consistent, it should be able to describe the isotropic-nematic (I-N) transition and nematic-smectic (N-S) equally well. An (almost) exactly solvable model for the I-N transition was developed by Onsager \cite{Onsager1949TheParticles}, and we might start there for inspiration. Onsager's approach relied on the virial expansion which, fortuitously, could be truncated. This is because, for highly anisotropic particles, the I-N transition happens at rather low concentrations. For the N-S transition this is not the case, and the virial expansion breaks down \cite{Kamien2014}. Hence, we must take a significantly different starting point for our model that can incorporate interactions between large numbers of molecules {\sl without appealing to the virial expansion}. 
		\begin{figure}\includegraphics[width=8cm]{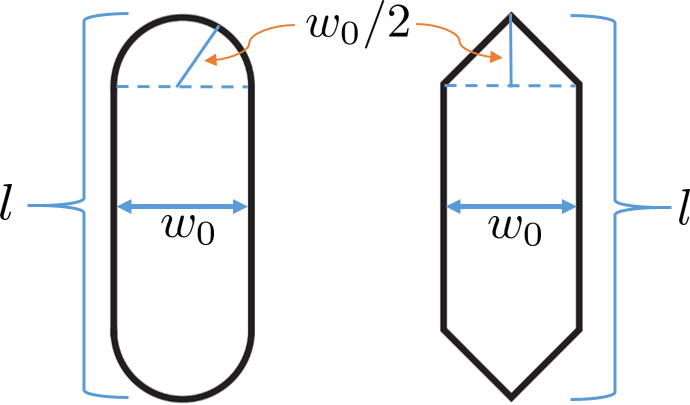}
		\caption{\label{fig:BK} Sketches of the particle shapes we consider. On the left is a ``bouba'', with a rectangular mid-section of width $w_0$ and semi-circular tips of radius $w_0/2$. The ``kiki'' is on the right, whose midsection is the same as the bouba, but whose tip is a triangle of height $w_0/2$. Both particles are of total length $\ell$.}
	\end{figure}

	Recall the bouba-kiki effect where, across cultures and languages, the word ``bouba'' is associated with rounded shapes and ``kiki'' with pointed shapes \cite{maluma,Ramachandran2001,Cwiek2022}\footnote{This effect was first realized by K{\"o}hler for shapes named ``maluma'' (rounded) and ``takete'' (pointed), although it is most famous now with the names bouba and kiki.}. We will argue that nature has a similar bias and expresses it by allowing boubas to form smectics more easily than kikis. 
To keep the model as simple as possible, we restrict our attention to two dimensions and simplify the particle structures.  We consider ``kikis'' instead of ellipsoids, and ``boubas'' instead of sphero-cylinders. Both the bouba and the kiki are of total length $l$, and have rectangular mid-sections with widths $w_0$, but their tips are different: The boubas have semi-circular tips, of radius $w_0/2$, whereas the kikis have triangular tips whose height is also $w_0/2$. These are sketched in Fig.(\ref{fig:BK}). 
	We model the N-CB molecules, with the same spirit of simplicity, as particles with a small body from which a flexible polymer tail of length $l_p$ emerges. For the cases of interest, 5-CB and 8-CB, the tail is relatively short, since it only includes a few repeating units. This makes the flexible polymer a crude model for the tail, as it assumes a very large number of monomers. Another simplifying but crude approximation we make is to ignore the size of the body, so that it has no excluded volume. Nevertheless this should not change the physics at the particle tips, which is our focus.
 
	Our approach is built on a simple construction of the free energy, which considers one test particle in a given background. By supposing that the dominant interaction between the particles is their excluded volume, we may understand the background as restricting the position of the test particle to a particular region. The size of this region controls the free energy. This allows interactions between large numbers of particles to be accounted for qualitatively in much the same way as successful tube theories in polymer physics \cite{Doi1986TheDynamics} or free volume theory \cite{Kirkwood1950}. We briefly outline this construction before showing how it is consistent with virial theory for a simple model of the I-N transition. We then apply it to the N-S transition for boubas and kikis and, subsequently, N-CB molecules. Our calculations demonstrate that boubas form smectics at lower densities than kikis, because the tip geometry  destabilizes the nematic phase. The same conclusion applies to the N-CB particles with long tails; (N+1)-CB makes a smectic at a lower density than N-CB.

	\label{sec:Construction}
	In this missive, we employ a general construction for the free energy that has been used before to determine the free energy of polymers subject to topological constraints \cite{Edwards1967StatisticalI,Edwards1967StatisticalII}: posit a test particle in state $\mathcal{T}$ placed in a background in the state $\mathcal{B}$. Later we will give specific examples of these states, for example one can imagine $\mathcal{T}$ to indicate if the test particle is in a ``nematic state'' or a ``smectic state'', for example. Assuming that the test particle is confined to a given region by the background allows us to determine the probability of realizing the test particle in some state, given the state of the background. We write this conditional probability as $P(\mathcal{T}|\mathcal{B})$. The probability of realizing the background state, $P(\mathcal{B})$ determines in what phase the system lies. For the purposes of our construction, we suppose it is known and is determined by minimizing the free energy. 
	
	We calculate the free energy of the system from Gibbs' definition
	\begin{equation}
		\beta F = \sum_{\mathcal{B}} \sum_{\mathcal{T}} P(\mathcal{T} \cap \mathcal{B}) \log P(\mathcal{T} \cap \mathcal{B})
	\end{equation}
	where $P(\mathcal{T} \cap \mathcal{B})$ is the probability of realizing $\mathcal{T}$ and $\mathcal{B}$, and the sums run over all possible states. Applying the identity $P(\mathcal{T} \cap \mathcal{B}) = P(\mathcal{B}) P(\mathcal{T}|\mathcal{B})$ and noting that $\sum_{\mathcal{T}} P(\mathcal{T}|\mathcal{B}) =1$ we find
	\begin{equation}
		\label{eqn:FEnGen}
		\beta F = \sum_{\mathcal{B}} P(\mathcal{B}) \log P(\mathcal{B}) + \sum_{\mathcal{B}, \mathcal{T}} P(\mathcal{B}) P(\mathcal{T}|\mathcal{B}) \log P(\mathcal{T}|\mathcal{B})
	\end{equation}
	The first term is understood as the free energy of the background, $\beta F_{\mathcal{B}}$, and the second as the free energy of a test particle in a given background, $\beta F_{\mathcal{T}}(\mathcal{B})$, averaged over all realizations of that background. The total free energy of the system is
	\begin{equation}
		\beta F = \beta F_\mathcal{B} + \left\langle \beta F_\mathcal{T} \right\rangle
	\end{equation}
with angle brackets denoting an average over the background. 

	\section{Isotropic-Nematic Transition}
	\label{sec:IN}
Let us demonstrate how this construction can be used to study liquid crystal transitions by applying it to the simplest model of the I-N transition \cite{Kamien2014}.  This involves a two-dimensional gas of rods (rectangles) which can only be oriented vertically or horizontally. The rods interact exclusively via their excluded volume, and it is supposed that each accesses every allowed position with equal probability. In this model, the isotropic phase is when the rods are vertical or horizontal with equal probability and the nematic when there is a bias one way or the other. Zwanzig studied, via a virial expansion, a three-dimensional version of this model where the rods can only point along the co\"ordinate axes \cite{zwanzig}; it can be specialized to two-dimensions where the analysis is relatively simple \cite{Kamien2014}. Here we demonstrate that our approach yields the same results as the more traditional approach but it also allows us to consider densities beyond which the virial expansion fails. 
	
The first step is to define the test particle and background states. The state of the test particle is determined by both its position and orientation, so we write $\mathcal{T} = (T, \textbf{r})$. Here $\textbf{r}$ is its position and the variable $T$ indicates if it is vertical ($V$) or horizontal ($H$). For the background, we suppose that every particle is in \textit{the same} orientation, given by the variable $B$. To completely specify the state, we then need to keep track of the positions of all the particles $\{\textbf{r}_i\}$ and we write $\mathcal{B} = (B , \{\textbf{r}_i\})$.

Next we need the conditional probability $P(\mathcal{T}|\mathcal{B})$. Given our assumptions, we have
\begin{equation}
	P(\mathcal{T}|\mathcal{B}) = \alpha_T \Theta_{TB}(\textbf{r},\{\textbf{r}_i\})
\end{equation} 
Here $\Theta_{TB}(\textbf{r},\{\textbf{r}_i\})$ is a unit indicator function which picks out the allowed positions $\textbf{r}$ of a test particle with orientation $T$ in a background of particles with orientation $B$ and positions $\{\textbf{r}_i\}$. The constant $\alpha_T$, which depends on the test particle orientation, is determined by ensuring $P(\mathcal{T}|\mathcal{B})$ is appropriately normalised. If the probability of the test particle being vertical is $p$, then
\begin{subequations}
		\begin{equation}
			P(V,\textbf{r} | B) = \frac{p}{\Omega_{VB}} \Theta_{V B}(\textbf{r},\{\textbf{r}_i\})
		\end{equation}
  and,
  		\begin{equation}
			P(H,\textbf{r} | B) = \frac{1-p}{\Omega_{HB}} \Theta_{H B}(\textbf{r},\{\textbf{r}_i\})
		\end{equation}
\end{subequations}
	for the two possible orientations and 
	\begin{equation}
		\label{eqn:accessiblevolumes}
		\Omega_{TB}(\{\textbf{r}_i\}) = \int d\textbf{r} \ \Theta_{TB}(\textbf{r},\{\textbf{r}_i\})
	\end{equation}
are normalization factors.
	Using these expressions we can directly compute $\beta F_{\mathcal{T}}(\mathcal{B})$ from (\ref{eqn:FEnGen})
		\begin{equation}
  \begin{split}
  		\beta F_{\mathcal{T}}(\mathcal{B}) &= \beta F_{0}(p) - p \log \Omega_{VB}(\{\textbf{r}_i\}) \\
    &- (1-p) \log \Omega_{HB}(\{\textbf{r}_i\})
  \end{split}
		\end{equation}
	where $\beta F_0(p) = p \log p + (1-p)\log (1-p)$ is the standard entropy of mixing.
	
	 What do we choose for $P(\mathcal{B})$? The state $\mathcal{B} = (B,\{\textbf{r}_i\})$ is realized with probability $P(\mathcal{B}) = \varphi(B) \psi(\{\textbf{r}_i\})$, with $\varphi$ being the orientational probability and $\psi$ the probability of the background particle positions. Both are taken to be independently normalized. Next we make the ``mean-field-like'' approximation to say that the probability of the background being vertical is \textit{the same} as that probability for the test particle, i.e. $\varphi(V) = p$. The same is of course true for the probability of being horizontal. Putting this into (\ref{eqn:FEnGen}) the total free energy as a function of $p$ is
	 \begin{equation}
	 	\label{eqn:INFEn1}
   \begin{split}
     \beta F(p)  &= 2 \beta F_0(p)  - p^2 \left\langle\log \Omega_{VV}\right\rangle - (1-p)^2 \left\langle\log \Omega_{HH}\right\rangle \\
     &- p(1-p) \left(\left\langle\log \Omega_{VH}\right\rangle+\left\langle\log \Omega_{VH}\right\rangle \right)
   \end{split}
	 \end{equation}
 	where $\Omega_{VV}$ is the accessible area to a vertical test particle in a vertical background, $\Omega_{VH}$ is that for vertical test particle in a horizontal background, and so forth. The angle brackets denote averaging over all positions of the background particles. This expression is simplified greatly by noting symmetries of the accessible areas, namely 
 	\begin{equation}
 		\label{eqn:INSymm}
 		\Omega_{VV} = \Omega_{HH} \equiv \Omega_{\parallel} \ \ \ \text{and} \ \ \ \Omega_{VH} = \Omega_{HV} \equiv \Omega_{\bot}
 	\end{equation}
	 It follows that the free energy is, up to a constant,
	 \begin{equation}
	 	\label{eqn:INFEn}
	 	\beta F(p) = 2 \beta F_0(p) - 2p(p-1) \left[\left\langle\log \Omega_{\parallel}\right\rangle- \left\langle\log \Omega_{\bot}\right\rangle\right]
 	 \end{equation}
  	Note the factor of two appearing in front of the entropy of mixing term, $\beta F_0$. This arises because, by artificially splitting the system into the test particle and the background, we are essentially considering \textit{two} separate populations of particles. As we shall see shortly, this factor of two is correct and leads to the same result as the virial approach. 
  	
  	To explore the I-N transition, we must find the equilibrium probability of the system being vertical, $p^*$, by minimizing $F(p)$:
  	\begin{equation}
  		\label{eqn:INPeqn}
  		\beta F'(p^*) = 0 = 2 \log \frac{p^*}{1-p^*} - 2(2p^* - 1) \Delta S
  	\end{equation}
  where  $\Delta S = \langle \log \Omega_{\parallel}\rangle
- \langle \log \Omega_{\bot}\rangle$. Evidently, when the two accessible areas, $\Omega_{\parallel}$ and $\Omega_{\bot}$, are both equal the only solution is $p^{*} = 1/2$. This is always a solution but, depending on $\Delta S$, this is not the minimum of the free energy. The difficult part of this approach is computing $\Delta S$ as a function of the density of the system. We will discuss this in more detail for the N-S transition but for now, guided by the knowledge that the I-N transition occurs at low density, we make a simple approximation valid in that limit. Namely, we employ free volume theory. The test particle may access the whole area of the system, $A$, except those parts where it overlaps with any background particle. For sufficiently low densities,  the background particles all independently exclude some area that does not depend on their position. Denoting this excluded area as $a_{\parallel,\bot}^{\text{exc}}$ in either the parallel or perpendicular case we may write, $\Omega_{\parallel,\bot} = A - N a^{\text{exc}}_{\parallel,\bot}$, and it follows that for small area density $\rho=N/A$:
\begin{equation}
	\Delta S = \log \left(\frac{1 - \rho a^{\text{exc}}_{\parallel}}{1 - \rho a^{\text{exc}}_{\bot}}\right) \approx \rho \left(a^{\text{exc}}_{\bot} - a^{\text{exc}}_{\parallel}\right)
\end{equation} 
Using this in (\ref{eqn:INFEn}) yields the same equation for $p^{*}$ as would be derived using Onsager's virial expansion approach \cite{Kamien2014}. This demonstrates the consistency of our construction with more traditional approaches for studying liquid crystal transitions. The advantage of our method is that the free energy is written in terms of the area accessible to a single particle. This is relatively straightforward to calculate (or estimate) even for concentrated systems where the virial expansion breaks down. As we shall see, this allows us to study the N-S transition in much the same way as the I-N transition.
	\section{Nematic-Smectic Transition}
	\label{sec:NS}
	An appealing aspect of our treatment of the I-N transition was that the continuous range of orientations a real particle can access was replaced by two discrete options; vertical and horizontal. To get this simplicity to carry over to the study of the smectic phase, we want to split the continuous range of positions into two distinct choices.
	
	The defining feature of the smectic phase is that the particles lie in distinct layers with a given separation. Let us say that these layers are all parallel to the $x$-axis and are separated by $h$. If our particles have total length $\ell$, then we must have $\ell < h < 2\ell$, for the layers to make sense. By analogy to the vertical-horizontal two state model of the I-N transition, let us suppose that there are two sets of such layers, ``solid'' and ``dashed''. The spacing between layers of the same type is $h$, but the layers are interleaved so that the distance between a solid and a dashed layer is $h/2$. The particles can be placed on either a solid or a dashed layer. Our goal is to find the free energy as a function of $p$, the probability that a particle occupies a solid layer, and to determine the equilibrium value $p^*$. When $p^* \neq 1/2$ we have a smectic-A phase, and we identify the state when $p^* = 1/2$ as the Nematic. Why should this be the case when there is still vertical layering?
	\begin{figure}\includegraphics[width=8cm]{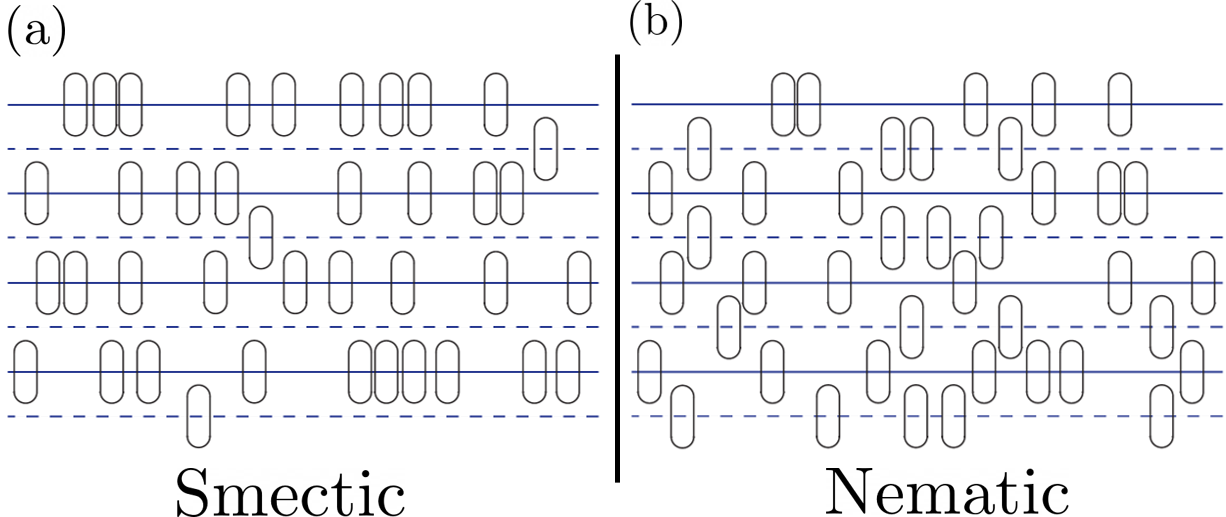}
		\caption{\label{fig:SmN} Sketches of the smectic and nematic phases in our model. In both panels the solid and dashed sets of lines are shown. In panel (a) the solid lines are preferred to the dashed by the particles, i.e $p \neq 1/2$. This is the smectic phase. Panel (b) has the solid and dashed lines occupied equally, $p = 1/2$. This is the nematic.}
	\end{figure}
	To see this, let us consider the definition of the smectic order parameter, $\mathcal{S}$ \cite{Frenkel1991}. The density of the particles as a function of $y$ can be expanded as a Fourier series
	\begin{equation}
		\rho(y) - \bar{\rho} = \sum_{n=1}^{\infty} \rho_n \cos\left(2\pi n y/h + \delta_n\right)
	\end{equation}
	where the $n=0$ mode defines the average density, $\bar{\rho}$, there is an arbitrary phase per mode, $\delta_n$, and $h$ is the aforementioned layer spacing. The coefficient of the $n=1$ mode defines the smectic order parameter, $\mathcal{S} \equiv \rho_1$. The nematic and smectic phases in this model are sketched in Fig.(\ref{fig:SmN}). When the solid and dashed layers are occupied with equal probability it is clear that
	\begin{equation}
		\rho(y) - \bar{\rho} = \rho_2 \cos\left(4\pi \frac{y}{h} +\delta_2 \right) + \cdots
	\end{equation}
	hence $\mathcal{S} = 0$ identically in this case. While there is now a new smectic with half the periodicity of the target phase, that is not the smectic for which we are looking! This is why we identify this as the nematic phase, even though there is a ``higher level'' layered order present. This situation is likewise true for the two-state model of the I-N transition: when vertical and horizontal orientations are equally likely, the nematic order parameter vanishes, but there is still 4-fold orientational order in the system.
	
	We construct the free energy as a function of $p$ using the same test particle and background construction as before. The state of the test particle, $\mathcal{T} = (T,x)$, tells us both whether it sits on a solid or dashed line and its $x$-position on that line and $T=S$ when it is on a solid line and $T=D$ when on a dashed line. We assume that all allowed $x$-positions of the test particle, not overlapping with a background particle, are equally likely.
	
\begin{figure}\includegraphics[width=8cm]{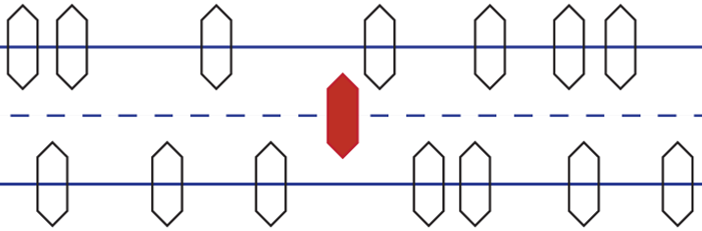}
		\caption{\label{fig:ExState} An example state of the test particle (picked out in red) and the background. Here, the test particle is in state $\mathcal{T} = (D,x)$, sitting on a dashed layer. The background is in state $\mathcal{B}=(S,\{x_i\})$, with all particles on solid layers. The set of co\"ordinates $\{x_i\}$ are the $x$-positions of the background particles. Only a selection of the background particles closest to the test particle are shown.}
	\end{figure}

	For the background state, $\mathcal{B}$, all of the particles occupy the same set of layers; either they are all on solid or all on dashed. We also need to keep track of the $x$-positions of all of the particles. This may appear intimidating, but notice that we need only keep track of those particles on layers which interact with the test particle, because all of the others will drop out of the calculation. We refer to the set of $x$-co\"ordinates for these particles by $\{x_i\}$, the range of the index $i$ depends on with how many layers the test particle interacts. Again $B = S$ for solid and $B=D$ for dashed. Furthermore, we may assume that each layer of the background has length $L$ and is  occupied by $N$ particles. We shall call the line density on each layer $\nu = N/L$. All together, we write $\mathcal{B} = (B,\{x_i\})$. In Fig.(\ref{fig:ExState}) we sketch an example state of the background and test particle. 

The conditional probability is
	\begin{equation}
		\label{eqn:SmecPTB}
		P(\mathcal{T}|\mathcal{B}) = \frac{p(T) \ \Theta_{TB}\left(x, \{x_i\}\right)}{\Omega_{TB}\left( \{x_i\}\right)}
	\end{equation}
	Here, $p(T)$ is the probability of the test particle being on a solid $T=S$ line or dashed $T=D$ lined and $\Theta_{TB}(x,\{x_i\})$ is a unit selector function picking out when the test particle at position $x$ does not overlap with any of the background particles. This latter function determines the ``accessible length'' for the test particle and provides the proper normalization
	\begin{equation}
		\Omega_{TB}\left(\{x_i\}\right) = \int_{-\infty}^{\infty} dx \ \Theta_{TB}\left(x,\{x_i\}\right)
	\end{equation}

	Define $p(T=S) = p$ for the probability of the test particle being on a solid line so $p(T = D) = 1 - p$. Applying the same mean-field  approximation as we did for the I-N transition we choose		$p(B = S) = p$ and $p(B = D) = 1 - p$
and follow the steps that led to (\ref{eqn:INFEn1}) to obtain
	\begin{equation}
 \begin{split}
				\beta F(p) = & 2 \beta F_0(p) - p^2 \left\langle \log \Omega_{SS} \right\rangle - (1-p)^2 \left\langle \log \Omega_{DD} \right\rangle\\
    &- p(1 -p) \left[\left\langle \log \Omega_{SD} \right\rangle+\left\langle \log \Omega_{SD} \right\rangle\right]
     \end{split}
	\end{equation}
	Similar symmetries to (\ref{eqn:INSymm}) apply due to the equivalence of shifting the whole system along $y$ by $h/2$ (solid/dashed duality);
	\begin{equation}
		\Omega_{SS} = \Omega_{DD} \equiv \Omega_{\text{o}}, \ \ \ \text{and} \ \ \ \Omega_{SD} = \Omega_{DS} \equiv \Omega_{\text{x}}.
	\end{equation}
	Up to a constant, the free energy is
		\begin{equation}
			\label{eqn:NSBKFen}
			\beta F(p) = 2 \beta F_0(p) -2p(p-1)\left[\left\langle\log \Omega_{\text{o}}\right\rangle-\left\langle\log \Omega_{\text{x}}\right\rangle\right].
	\end{equation}
	Note how similar this is in structure to (\ref{eqn:INFEn}) for the I-N transition. Hence, the equation determining $p^*$ is precisely the same as (\ref{eqn:INPeqn}):
	\begin{equation}
		\label{eqn:phieq}
		\log\frac{p^*}{1-p^*} = (2p^* - 1) \Delta S
	\end{equation}
	where we have defined $\Delta S = \left\langle\log \Omega_{\text{o}}\right\rangle-\left\langle\log \Omega_{\text{x}}\right\rangle$. We see that when $\Delta S > 2$, a smectic phase forms with $p^* \neq 1/2$. So the problem all comes down to computing $\Delta S$ for the boubas and kikis and the N-CBs -- the key here is that we do not need to rely upon the low-density limit.  In the following we will estimate $\Delta S$ directly in the spirit of the Tonks gas \cite{Tonks1936TheSpheres}.  Note that $\Delta S$ is a function of the layer spacing, $h$, the density on each layer $\nu$ and the average density $\bar{\rho}=\text{Number}/\text{Area} = N/(Lh) = \nu/h$. 	
	Our aim is to show that boubas undergo a N-S transition at a \textit{lower} density than kikis, and to elucidate the difference that the tip shape makes. For the N-CBs, we would like to show that the \textit{larger} N is, the lower the density at which the smectic forms. We do not aim to precisely determine the phase boundary in any case, that would require a more sophisticated method.
	
	\subsection{Boubas versus Kikis}
\begin{figure}\includegraphics[width=8cm]{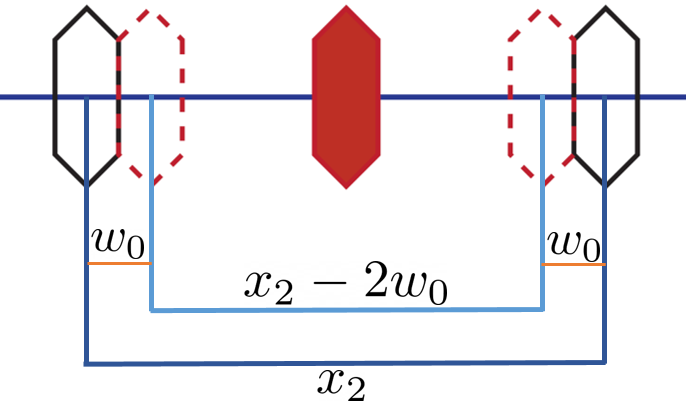}
		\caption{\label{fig:Dot} A sketch of the test particle (in red) on a solid layer, when the background particles are also all on solid layers. The two background particles closest to the test particle are indicated. These two are separated by a distance $x_2$. Each excludes a length of $w_0$ to the test particle, so that the accessible length to it in this configuration is $\Omega_{\text{o}} = x_2 - 2 w_0$.}
	\end{figure}

	The whole calculation boils down to computing $\langle \log \Omega_{\text{o}} \rangle$ and $\langle \log \Omega_{\text{x}} \rangle$. In the first case, the test particle only interacts with those background particles on its own layer, because of the restriction $h < 2\ell$. This also means that the result will be identical for boubas and kikis, because the tip geometry is irrelevant when interacting with mesogens on the {\sl same} layer. The starting point is an expression for $\Omega_{\text{o}}$. Let $x_2$ be the distance between the centers of the closest background particle to the left and right of the test particle. The accessible length is then simply
	\begin{equation}
		\Omega_{\text{o}} = x_2 - 2 w_0,
	\end{equation}
	because each background particle excludes a length $w_0$, as shown in Fig.(\ref{fig:Dot}). So, we must compute
	\begin{equation}
		\label{eqn:avgo}
		\langle \log \Omega_{\text{o}} \rangle = \int dx_2 \ P(x_2) \log(x_2 - 2 w_0),
	\end{equation}
	where $P(x_2)$ is the probability of realizing the distance $x_2$. Each layer is a Tonks gas \cite{Tonks1936TheSpheres}, a one dimensional gas of finite sized particles interacting only via excluded volume. The distance $x_2$ is the next-nearest-neighbor distance for such a gas, and its distribution, $P(x_2)$ was calculated by Tonks. This allows us to explicitly calculate (\ref{eqn:avgo}). This is done in Appendix \ref{app:BK}, but here we make an approximation which make our analysis very simple, but does not change the outcome. The approximation replaces
	\begin{equation}
	\langle \log \Omega_{\text{o}}\rangle \to \log \langle \Omega_{\text{o}}\rangle = \log \left( 2/\nu - 2 w_0 \right), 
	\end{equation}
	where we have used Tonks' result $\langle x_2\rangle = 2/\nu$. 
	
\begin{figure}\includegraphics[width=8cm]{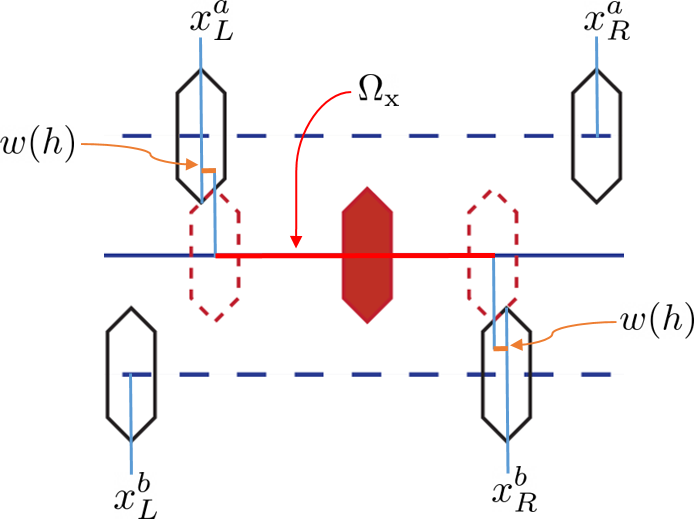}
	\caption{\label{fig:Cross} The red test particle sits on a solid layer in a background of particles on dashed layers. The four closest background particles to the test particle are shown; two on the layer above and two on the layer below. Using the conventions of equation (\ref{eqn:OmegaX}), this is the configuration $(baba)$. Because of the shape of the tips, the background particles exclude a length of $w(h) < w_0$. The length accessible to the test particle is $\Omega_{\text{x}}$, as given in (\ref{eqn:OmegaXbaba}).}
	\end{figure}

	Now we turn our attention to $\langle \log \Omega_{\text{x}}\rangle$. Once again we shall replace this with $\log \langle \Omega_{\text{x}} \rangle$, but the complete calculation is in Appendix \ref{app:BK}. In this case, there are no background particles on the same layer as the test particle. However, the occupied layer above is only vertically separated from it by $h/2$, so it may interact with that layer and it likewise interacts with the layer beneath. Let us refer to the closest background particles on the left and right as $x_L$ and  $x_R$, respectively. We supply these with the superscripts $a$ or $b$ to indicate if they come from the layer above or below the test particle so that, $x_L^{a}$ is the position of the closest particle on the layer above the test particle to its left and so on. Now, we can write $\Omega_{\text{x}}$ as
	\begin{equation}
		\Omega_{\text{x}} = \min_{i \in (a,b)} x^j_{R} - \max_{i \in (a,b)} x^{i}_{L} - 2 w(h)
	\end{equation}
	so that the absolute left and right limits for the test particle are set by the background particles closest to it. The function $w(h)$ is the length excluded by the particle, its effective width, which must be a function of $h$ because of the shape of the tip. Note that the function $w(h)$ is different for different tip shapes. This expression requires us to consider the four possible arrangements of background particles. One example is for the closest on the left to come from the layer above and that on the right to come from the layer below. In this situation if we move from all the way to the left to all the way to the right, we encounter the background particles from different layers in the order; below, above, below, above. This situation is sketched in Fig.(\ref{fig:Cross}). We shall refer to this configuration as $(baba)$, and all others accordingly. The accessible lengths in each case are simply 
	\begin{subequations}
		\label{eqn:OmegaX}
	\begin{equation}
	(abab) \to \Omega_{x} = x_R^{a} - x_L^{b} -2w(h),
	\end{equation}
	\begin{equation}
	(abba) \to \Omega_{x} = x_R^{b} - x_L^{b}- 2w(h),
	\end{equation}
	\begin{equation}
	(baab) \to \Omega_{x} = x_R^{a} - x_L^{a}-2w(h),
	\end{equation}
	\begin{equation}
		\label{eqn:OmegaXbaba}
	(baba) \to \Omega_{x} = x_R^{b} - x_L^{a}-2w(h).
	\end{equation}
	\end{subequations}

By symmetry, all four of these situations are realized with equal probability, so that the average $\langle \Omega_{\text{x}}\rangle$ over all realizations of the background is 
		\begin{equation}
			\langle  \Omega_{\text{x}}\rangle = \frac{1}{4} \left[2\langle x_R^a - x_L^a\rangle + 2\langle x_R^b - x_L^b\rangle - 8 w(h)\right]
		\end{equation}
The angle brackets here denote averaging over all positions $x^{a,b}_{R,L}$. Notice that the combinations $x_R^{a,b}- x_L^{a,b}$ are both the nearest neighbor distance in the Tonks gas, $x_1$. The average of this is, $\langle x_1 \rangle = 1 / \nu$ so that
	\begin{equation}
	\log \langle \Omega_{\text{x}}\rangle = \log\left( 1/\nu - 2 w(h)\right)
	\end{equation}
We now have an expression for $\Delta S$, and the condition for a Smectic phase is
	\begin{equation}
		\Delta S = \log 2 + \log\left(\frac{1-\nu w_0}{1-2\nu w(h)}\right) > 2
	\end{equation}
This can be cast as a condition on $w(h)$
	\begin{equation}
		\label{eqn:wcondmain}
		2w(h) > \frac{2}{e^2} w_0 + \frac{1}{2 \nu}(e^2 -2)
	\end{equation}
or, assuming that $\nu$ is relatively large, a looser condition is $2 w(h) \gtrsim w_0$. This is the result of the more detailed analysis in Appendix \ref{app:BK} and is understood simply as comparing the length excluded to the test particle by the background particles, $2 w(h)$, to that excluded by the background to themselves, $w_0$. Crudely speaking, does the background allow enough room for the test particle to muscle its way in between the layers? Naturally, this will depend on the width of the particle's shoulders expressed through its tip geometry. This is quantified by understanding the function $w(h)$.
	
\begin{figure}\includegraphics[width=8cm]{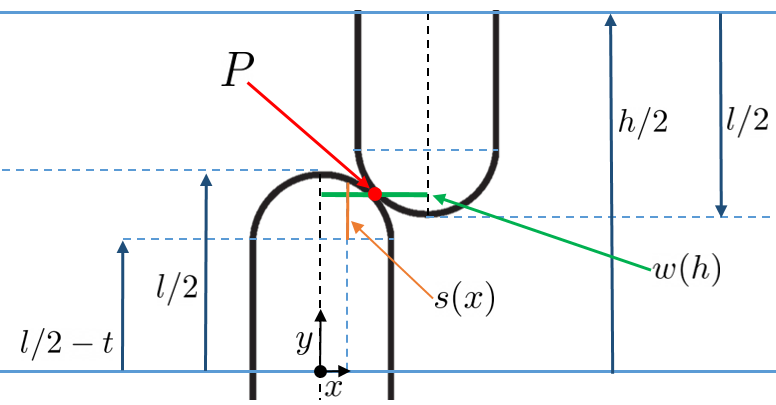}
		\caption{\label{fig:Kissing} A sketch of two particles on layers separated by $h/2$ colliding at their tips. The particles shown are boubas, but the geometry is equivalent for any shape. The symmetric tip shape function, $s(x)$, is indicated in orange. The distance between the centers of the two particles is shown in green; this is the excluded length, $w(h)$. Equation (\ref{eqn:wh}) for $w(h)$ is found by considering the $y$-co\"ordinate of the point $P$ where the particles meet.}
	\end{figure}

Consider a generic particle of width $w_0$ whose tip has a symmetric shape described by the function $y=s(x)$. This function describes the height of the tip above the midsection of the particle at a position $x$ along its width. We require $-w_0/2 \leq x \leq w_0/2$, and symmetry enforces $s(x) = s(-x)$. We suppose that the full length of the particle is $\ell$ and that the total length of one tip is $t$. The function $w(h)$ is determined by finding the point $P$, indicated in Fig.(\ref{fig:Kissing}), where two oppositely oriented particle tips touch if the centers of the particles are vertically separated by a distance $h/2$. Considering only the lower particle we have
	\begin{equation}
		P = \bigg(w(h)/2, \ \ell/2 - t + s\left(w(h)/2\right)\bigg)
	\end{equation}
	and considering the upper particle we find,
	\begin{equation}
		P = \bigg(w(h)/2, \ h/2 - \ell/2 + t - s\left(-w(h)/2\right)\bigg)
	\end{equation}
	These expressions must both represent the same point, hence 
	\begin{equation}
		\label{eqn:wh}
		2s\left(\frac{w(h)}{2}\right) = \frac{h}{2} - \ell + 2t.
	\end{equation}
	If we know the function $s(x)$ describing the tip shape, then we can find $w(h)$. For boubas and kikis, $s(x)$ is particularly simple. 

	A bouba has a semi-circular tip of radius $w_0/2$ so $t = w_0/2$ and 
		$s_B(x)=\sqrt{\left(\frac{w_0}{2}\right)^2 - x^2}$,
	which leads to 
		$w_B(h) = \sqrt{w_0^2 - (h/2 - \ell + w_0)^2}$.
	For kikis, whose tips are triangular with height $t = w_0/2$ and so
		$s_K(x) = \frac{w_0}{2} - |x|$,
and hence 
		$w_K(h) = \ell - \frac{h}{2}$.

	With the condition (\ref{eqn:wcond}) along with the functions $w_B(h)$ and $w_K(h)$ we can find conditions for which values of $h$ boubas and kikis form smectics. For boubas 
	\begin{equation}
		h_B \leq 2l - (2 - \sqrt{3}) w_0 \approx 2\ell - 0.27 w_0
	\end{equation}
	and for kikis
	\begin{equation}
		h_K \leq 2\ell - w_0
	\end{equation}
	Evidently, boubas will form a smectic for a larger layer spacing $h$ than kikis. Because we can relate $h$ to the number density $h = \nu/\bar{\rho}$, this implies that boubas make a smectic at a \textit{smaller} average density $\bar{\rho}$ than kikis.   It is essential to note that the entropy difference arises from considering test rods that are {\sl not} on the background smectic layer.  In this sense, it is the nematic phase that is being changed, not the smectic.  When the tips are pointier there is more opportunity for a rod to find space in half layer between the smectic layers.
 
    It is interesting to consider briefly the limiting case when the particle tips become flat. Now the particles are rectangles with dimensions $w_0 \times \ell$. The effective width for these shapes has a step; $w(h)= 0$ for $h\geq2\ell$ and $w(h) = w_0$ for $h<2\ell$. The calculation given above tells us that these rectangles form a smectic when the layer spacing becomes $h < 2\ell$. However, applying Frenkel's rescaling argument \cite{Frenkel1991}, we can map the rectangles onto a system of $w_0 \times w_0$ squares. We would then say that these squares form a smecticas soon as $h < 2 w_0$. Nothing prevents this from happening in principle but such a phase is not observed in simulations \cite{Ree1972,Wojciechowski2004}. Though some calculations do predict a smectic phase, it is expected to be unstable to fluctuations for infinite systems \cite{Belli2012}. In our case, when the layer spacing is just larger than the transition value $2 w_0$, the system should be ``nematic'' with the dashed and solid layers equally occupied. Given that these layers are spaced by a little more than $w_0$, the squares will be just touching those on the layer above or below. In this way, the order in the $y$-direction is \textit{the same} as would be observed in a crystal but the difference between this state and a crystal is the order in the $x$-direction where we have a Tonks gas. It could be argued that the instability shown by our calculation when the layer spacing is decreased is actually the instability to forming the crystal. Given that the particles can only occupy layers separated by $h/2$ and $h$, this instability will artificially give rise to a smectic phase for squares.  
	\subsection{N-CB Molecules} 
	\label{sec:Poly}
	Finally, let us consider the N-CB molecules. We use the same free energy construction as before for the boubas and kikis. This time, we must also keep track of the degrees of freedom for the test particle and background polymer tails. For simplicity we ignore the size of the body of the molecule and the self-excluded volume of tail. We are lead to exactly the same form of equation for $p^{*}$ as (\ref{eqn:phieq}), and exactly the same condition for the smectic phase, namely,
	\begin{equation}
		\Delta S^{\text{poly}} \equiv \left\langle \log \Omega^{\text{poly}}_{\text{o}} \right\rangle - \left\langle \log \Omega^{\text{poly}}_{\text{x}} \right\rangle \geq 2. 
	\end{equation}
	Here $\log \Omega^{\text{poly}}_{\text{o}}$ is the entropy of the polymer tail of the test particle when it sits on a solid line in a background of particles on solid lines, and $\log \Omega^{\text{poly}}_{\text{x}}$ is the entropy when the test particle is on a dashed (solid) line and the background particles are on solid (dashed) lines. In this expression, the angle brackets denote averaging over all positions of the background particle bodies and all configurations of their polymer tails. Just as for the boubas and kikis, we assume that the particle density on each layer is $\nu$.
	
	To make progress, we make the same approximation as before $	\left\langle \log \Omega^{\text{poly}} \right\rangle \approx \log \left\langle  \Omega^{\text{poly}} \right\rangle$.
	In this way, each term can be understood as the entropy of the test polymer tail in a fixed average background. Due to the excluded volume of the background polymer tails, the presence of the background acts to restrict the accessible configurations of test polymer. A simple model for this is to say that the test polymer is confined to a rectangular box with dimensions $L^x \times L^y$. The lengths $L^{x,y}$ depend on whether we consider $\left\langle  \Omega^{\text{poly}}_{\text{o}} \right\rangle$ or $\left\langle  \Omega^{\text{poly}}_{\text{x}} \right\rangle$. 
	
	In the former case, the width in the $x$-direction is the average next-to-nearest neighbor distance in the Tonks gas, $L_{\text{o}}^x = 2/\nu$. The height in the $y$-direction in this case is the distance between the two closest layers to that on which the test particle sits, $L_{\text{o}}^y = 2 h$. In the latter case, the width and heights are halved. The width is the \textit{nearest} neighbour distance in the Tonks gas $L_{\text{x}}^x = 1/\nu$, and, if the test particle is on a dashed (solid) layer, the height is the distance between the two closest solid (dashed) layers $L_{\text{x}}^y = h$.
	
	It is now a straightforward polymer physics problem \cite{Doi1986TheDynamics, Edwards1969TheI} to compute the entropies of the polymers in these boxes. While we can obtain expressions of $\left\langle  \Omega^{\text{poly}}_{\text{o}} \right\rangle$ and $\left\langle  \Omega^{\text{poly}}_{\text{x}} \right\rangle$ for any polymer chain length $l_p$ (see Appendix \ref{app:NCB}), let us focus for now on two important limits; polymers much smaller than the boxes, and those much longer. In the first instance we must have $l_p \ll h,\nu^{-1}$ and we find 
		\begin{equation}
			\label{eqn:OmegaShort}
			\left\langle\Omega_{\text{o}}\right\rangle \sim \frac{2}{\nu} + \mathcal{O}(l_p/h), \ \ \ \text{and} \ \ \ \left\langle\Omega_{\text{x}}\right\rangle \sim \frac{1}{\nu} + \mathcal{O}(l_p/h).
		\end{equation}
	Here, there is no smectic transition since $\Delta S \approx \log 2 < 2$. 
	
	In the second case, where the polymers are long, we must have $l_p \gg h, \nu^{-1}$. This leads to 
 \begin{subequations}
 \label{eqn:OmegaLong}
  		\begin{equation}
			\left\langle\Omega_{\text{o}}\right\rangle \sim \frac{2^6}{\pi^3 \nu} e^{-l_p^2\left(\nu^2 + h^{-2}\right)/4},
		\end{equation} 
  and,
  \begin{equation}
			\left\langle\Omega_{\text{x}}\right\rangle \sim \frac{2^5}{\pi^3 \nu} e^{-l_p^2\left(\nu^2 + h^{-2}\right)}.
		\end{equation} 
 \end{subequations}
	Therefore the smectic condition is
	\begin{equation}
		\Delta S = \log 2 + \frac{3}{4}l_p^2 \left(\frac{1}{h^2}+ \nu^2\right) \geq 2. 
	\end{equation}
	In the same way as for the boubas and kikis, this can be read as a condition on the layer spacing, $h$. Namely, for a smectic, we must have
	\begin{equation}
		h^2 \leq \left(\frac{4}{3 l_p^2} (2 - \log 2) - \nu^2 \right)^{-1} \sim l_p^2
	\end{equation}
	So it follows that particles with longer polymer tails form a smectic at \textit{larger} layer spacings than those with shorter tails. This implies that they also form at \textit{lower} densities. The limit of very short polymer tails also showed us that there are some tails which are so short that they do not form smectics at all. The physical reason for these differences is essentially the same as that for the boubas and kikis; the longer polymer tails make it harder for particles to penetrate between the smectic layers. 
			\begin{figure}\includegraphics[width=8cm]{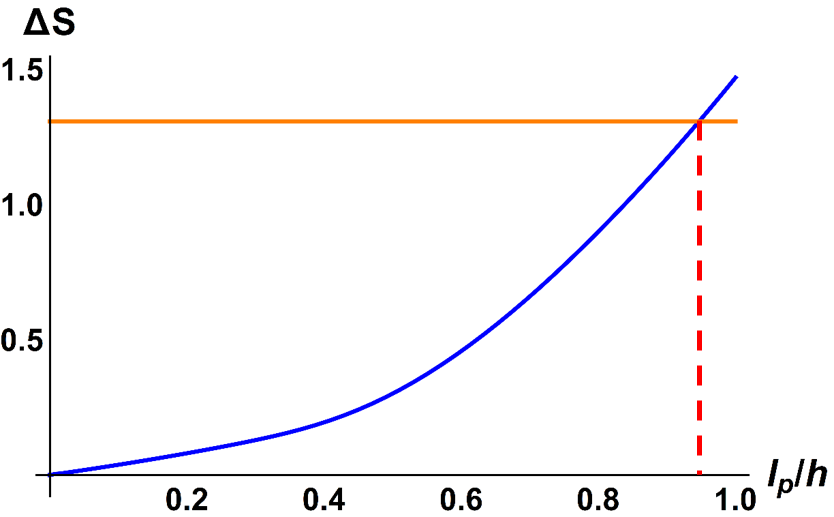}
		\caption{\label{fig:DeltaS} $\Delta S$ for plotted a function of the ratio of the polymer tail length to the layer spacing, $l_p/h$. The blue curve is $\Delta S$ and the orange line is the value it must exceed for a smectic to form. This happens for $l_p/h$ indicated by the red dashed line. This critical ratio is less than, but close to, unity.}
	\end{figure}
	
	We can also plot the full form of $\Delta S$ as a function of $l_p/h$ at fixed density, assuming that $\bar{\rho} = h^{-2}$. This is shown in Fig.(\ref{fig:DeltaS}). There we see that the smectic condition is met for longer polymers, with values of $l_p/h \lesssim 1$.  
	
	At this point one might raise concern about our choice of box size. While the widths in the $x$-direction are clear enough, there may be some question about the chosen heights. The background may be thought of as layers of polymer brushes of some height $H < h$. It is intuitive to expect the these brushes prevent the test polymer from reaching all the way to the nearest layer, by virtue of the excluded volume interactions. To capture this effect, the box height should be reduced by an amount proportional to the brush height; $h \to h - \alpha H$, where $\alpha < 1$. The brush height depends on $l_p$ and $\nu$ and, with reference to the simple arguments of Alexander \cite{Alexander1977} and de Gennes \cite{DeGennes1976, DeGennes1980}, as well as the more sophisticated results of Milner, Witten, and Cates \cite{Milner1988TheoryBrush}, it must increase when $l_p$ or $\nu$ are increased. This modification only serves to make shorter polymer tails worse at making smectics compared to longer tails. While more involved treatments of the polymer tail entropy are possible and will alter the details of our conclusions, we do not expect them to change the underlying result that, longer polymer tails de-stabilize the nematic phase by making the interstices between layers less accessible.
	\section{Conclusions}
	\label{sec:Concs}
	We have explored which particles can form a smectic-A phase by means of a simple two dimensional model. In this model, we consider a single test particle in a fixed background which restricts the positions of the test particle to a well defined region. The size of the region determines the entropy of the test particle and, by means of a mean-field-like approximation, the free energy of the system. This construction qualitatively includes the interactions between a large number of particles allowing it to be applied to higher density systems for which approaches based on the virial expansion are not valid. In particular this allows the nematic-smectic transition to be treated on the same footing as the isotropic-nematic. We demonstrated that our construction is exactly consistent with virial approaches to the I-N transition in the low density limit. 
	
	We considered the N-S transition for two different rigid particle shapes and for N-CB molecules. The rigid particles chosen were boubas and kikis, shown in Fig.(\ref{fig:BK}). These model three dimensional sphero-cylinders and ellipsoids respectively. It has been noted previously that ellipsoids \textit{do not} form a smectic but sphero-cylinders do. Similarly it is known that 8-CB forms a smectic while 5-CB does not. Our model for these molecules is a small body with a polymer tail of a given length. It is expected then that longer polymer tails lead to smectics at lower densities. 
	
	The analysis of our simple model shows that particles with ``fatter'' tips form smectics at lower densities than those with ``thinner'' ones. The reason for this is that fatter tips allow less space between the smectic layers to any rogue interloper trying to make a new home away from its own layer, thereby de-stabilizing the nematic at a given density. This same reasoning applies to the N-CB molecules, where it is the longer polymer tails which make the region between the smectic layers less accessible.
	
	Of course the approach that we have taken is only approximate and will not give accurate predictions for the phase boundary. In the same way, we have not addressed the smectic-crystal transition. This would complete the picture by demonstrating that for kikis, say, the N-S transition actually happens at a \textit{higher} density than crystallization, but this is beyond the reach of our simple model. Due to the reduction of degrees of freedom in two dimensions, the predicted order of the phase transitions discussed may be incorrect. In principle our approach may be followed in 3D, but this could result in sufficiently complicated analyses that our sacrifices made in the name of simplicity may not be worthwhile. Nevertheless, our simple arguments elucidate the physics governing which particles can form smectic phases.
	
This work was supported by a Simons Investigator grant from the Simons Foundation to R.D.K.
	
	\appendix
	\section{Boubas and Kikis}
	\label{app:BK}
	Here we compute $\Delta S$, from equations (\ref{eqn:NSBKFen}) and (\ref{eqn:phieq}), relevant for the N-S transition of boubas and kikis \textit{without} the approximation $\langle \log \Omega \rangle \approx \log \langle \Omega\rangle$.
	
	The first step is computing $\langle \log \Omega_{\text{o}} \rangle$. This is given in equation (\ref{eqn:avgo}) in terms of $P(x_2)$, the distribution of next-nearest neighbour distance in the Tonks gas. This distribution may be found exactly \cite{Tonks1936TheSpheres}, and is given by
		\begin{equation}
		P(x_2) = \frac{\nu^2 \left(x_2 - 2 w_0\right)}{(1- \nu w_0)^2}  \exp\left(- \frac{\nu}{1 - \nu w_0} \left(x_2 - 2 w_0\right)\right)
	\end{equation}
	for $x_2 \geq 2 w_0$ and zero otherwise. Integrating, we have
	\begin{equation}
		\langle \log \Omega_{\text{o}}\rangle = 1 - \gamma + \log \frac{1 - \nu w_0}{\nu},
	\end{equation}
	where $\gamma$ being the Euler-Mascheroni constant \cite{Abramowitz1964HandbookTables}.
	
Next we require $\langle \log \Omega_{\text{x}} \rangle$. As discussed in the main text, we need to consider the four cases (\ref{eqn:OmegaX}). It is convenient for us to write these positions in terms of $x_1^{i} = x_L^i - x_R^i$, that is the nearest-neighbur distance in the Tonks gas in layer $i$. It is also useful to introduce the separation of the closest particles on the left, $\Delta_L = x_L^{a}  - x_L^b$. Note that, in order for ``left'' and ``right'' to make sense, we must have $|\Delta_L| \leq x_1^{i}$. This gives us
\begin{subequations}
	\begin{equation}
		(abab) \to \Omega_{x} = x_1^a - |\Delta_L|-2w(h)
	\end{equation}
	\begin{equation}
		(abba) \to \Omega_{x} = x_1^b-2w(h)
	\end{equation}
	\begin{equation}
		(baab) \to \Omega_{x} = x_1^a-2w(h)
	\end{equation}
	\begin{equation}
		(baba) \to \Omega_{x} = x_1^b - |\Delta_L|-2w(h)
	\end{equation}
\end{subequations}
\widetext
 All of these are realized with equal probability, so that the averaging $\langle \log \Omega_{\text{x}}\rangle$ over all realizations of the background results in
\begin{equation}
	\begin{split}
		\langle \log \Omega_{\text{x}}\rangle = \frac{1}{4}\int dx_1^{a} P(x_1^a) \int dx_1^b P(x_1^b) &\int d \Delta_L P(\Delta_L) \bigg[\log (x_1^a-2w(h)) + \log (x_1^b-2w(h)) \\
		&+ \log (x_1^a - |\Delta_L|-2w(h)) + \log (x_1^b - |\Delta_L|-2w(h))\bigg]
	\end{split}
\end{equation}
Notice that the distributions $P(x_1^a)$ and $P(x_1^b)$ are \textit{the same} and normalized, so that first two terms in the square brackets are the same as are the final pair. This leaves
\begin{equation}
	\langle \log \Omega_{\text{x}}\rangle = \frac{1}{2}\int dx_1 P(x_1)\int d \Delta_L P(\Delta_L) \bigg[\log (x_1-2w(h)) +\log (x_1 - |\Delta_L|-2w(h))\bigg]
\end{equation}
To take the integral over $\Delta_L$ we need its probability distribution. Because the $a$ layer and $b$ layer are independent of each other this must be uniform. The only restriction is on its magnitude $|\Delta_L| \leq x_1$. Hence,
\begin{equation}
	\langle \log \Omega_{\text{x}}\rangle = \frac{1}{2}\int dx_1 P(x_1)\int_{0}^{x_1} \frac{d \Delta_L}{x_1} \bigg[\log (x_1-2w(h)) +\log (x_1 - |\Delta_L|-2w(h))\bigg]
\end{equation} 
\endwidetext
and so
\begin{equation}
	\label{eqn:avgx}
	\langle \log \Omega_{\text{x}}\rangle = -\frac{1}{2}+\int dx_1 P(x_1)\int_{0}^{x_1} \log (x_1-2w(h)).
\end{equation}
This is now written in an analogous way with (\ref{eqn:avgo}), only now in terms of the distribution of nearest-neighbor separations in a Tonks gas $P(x_1)$. This distribution was also worked out by Tonks \cite{Tonks1936TheSpheres}
\begin{equation}
	P(x_1) = \frac{\nu}{1-\nu w_0} \exp\left(-\frac{\nu}{1-\nu w_0} \left(x_1 - w_0\right)\right)
\end{equation}
This is straightforward, although this time the result is not quite as compact,
\begin{equation}
	\langle\log \Omega_{\text{x}}\rangle = -\frac{1}{2}+\log\frac{1-\nu w_0}{\nu} + \int_{0}^{\infty} d\xi \ e^{-\xi} \log(\xi + \alpha)
\end{equation}
with $\alpha = \nu (w_0 - 2 w(h))/(1 - \nu w_0)$. While the $\xi$ integral can be written in terms of incomplete Gamma functions \cite{Abramowitz1964HandbookTables} it is not particularly illuminating. 

Now we have $\Delta S$, and the condition for a stable smectic phase is
\begin{equation}
	\Delta S = \frac{3}{2} - \gamma - \int_{0}^{\infty} d\xi \ e^{-\xi} \log(\xi + \alpha)> 2
\end{equation}
The parameter $\alpha$ is a function of both the tip shape, and the density. Therefore, this inequality relates the density for the N-S transition to the tip shape. When the integral in this inequality becomes sufficiently negative, the inequality is satisfied. The integral is positive for all positive $\alpha$, but becomes infinitely negative when $\alpha < 0$. Thus, given $\nu \geq 0$ and $w_0 \geq 0$, the condition required for the smectic phase is,
\begin{equation}
	\label{eqn:wcond}
	2 w(h) \geq w_0
\end{equation}
This is qualitatively the same as the relation (\ref{eqn:wcondmain}) derived using the approximations in the main text.

\section{N-CB Molecules}
\label{app:NCB}
Here we compute $\Delta S$ for the N-S transition of N-CB molecules. The approximation $\langle \log \Omega\rangle \approx \log \langle \Omega\rangle$ is required here to avoid a complicated self-consistent treatment of the polymer. Within this approximation, each term in $\Delta S$ can be thought of as the entropy of a polymer in a 2D box with dimensions $L_x \times L_y$. Finding this entropy is a standard problem \cite{Doi1986TheDynamics} and the starting point is the polymer Green's function $G(x,x';y,y'| n)$ which solves
\begin{equation}
\begin{split}
    	\left[\frac{\partial}{\partial n} - \frac{b^2}{6}\left(\frac{\partial^2}{\partial x^2}+\frac{\partial^2}{\partial y^2}\right)\right] &G(x,x';y,y'|n) \\
     &= \delta(x-x') \delta(y-y') \delta(n),
\end{split}
\end{equation}
and is subject to the boundary conditions at the walls of the box
\begin{equation}
	G(x=0,L_x, x' ; y,y' | n ) = G(x , x' ; |y|=L_y/2,y' | n ) = 0.
\end{equation}
Here the co\"ordinates $x'$ and $y'$ represent the horizontal and vertical positions of the start of the polymer chain. Note that $x'$ may take any value allowed by the box, but we require $y'=0$. The variable $n$ represents the number of monomers making up the chain and $b$ measures the bond lengths between monomers. The entropy can be computed via
\widetext
\begin{equation}
	\label{eqn:OmegaApp}
     	\Omega(L_x,L_y) = \int_{0}^{L_x} dx\int_{0}^{L_x} dx' \int_{-L_y/2}^{L_y/2} dy \ G(x,x';y,y'=0|n).
\end{equation}
The Green's function is found by separation of variables $G = g_x(x,x'|n) g_y(y|n)$, with 
\begin{subequations}
	\begin{equation}
		g_x(x,x'|n) = \frac{2}{L_x} \sum_{m=1}^{\infty} \sin\left(\frac{m \pi x}{L_x}\right) \sin\left(\frac{m \pi x'}{L_x}\right) \exp\left(- m^2 \frac{\pi^2  n b^2}{6 L_x^2}\right),
	\end{equation}
and 
\begin{equation}
	g_y(y|n) = \frac{2}{L_y} \sum_{m=0}^{\infty} \cos\left(\frac{(2m+1) \pi y}{L_y}\right) \exp\left(- (2m+1)^2 \frac{\pi^2  n b^2}{6 L_y^2}\right).
\end{equation}
\end{subequations}
Identifying the length of the polymer chain as $l_p^2 = \pi^2 n b^2/6$ and taking the integrals in (\ref{eqn:OmegaApp}) we find
\begin{equation}
	\Omega(L_x,L_y) = \frac{2^5}{\pi^3} L_x\sum_{p \in \text{Odd}}\sum_{m=0}^{\infty} \frac{(-1)^m}{p^2 (2m+1)} \exp\left[- l_p^2 \left(\frac{p^2}{L_x^2} + \frac{(2m+1)^2}{L_y^2}\right)\right].
\end{equation}
Taking the limit that the polymer is much smaller than the box, $l_p \ll L_x,L_y$ yields
\begin{equation}
	\Omega(L_x,L_y) \sim  L_x,
\end{equation}
For the opposite limit $l_p \gg L_x,L_y$ we find
\begin{equation}
	\Omega(L_x,L_y) \sim \frac{2^5}{\pi^3} L_x \exp\left[-l_p\left(\frac{1}{L_x^2} + \frac{1}{L_y^2}\right)\right].
\end{equation}
These expressions reduce to (\ref{eqn:OmegaShort}) and (\ref{eqn:OmegaLong}) of the main text when the appropriate box dimensions are used. 
\endwidetext
%


\begin{thebibliography}{33}%
\makeatletter
\providecommand \@ifxundefined [1]{%
 \@ifx{#1\undefined}
}%
\providecommand \@ifnum [1]{%
 \ifnum #1\expandafter \@firstoftwo
 \else \expandafter \@secondoftwo
 \fi
}%
\providecommand \@ifx [1]{%
 \ifx #1\expandafter \@firstoftwo
 \else \expandafter \@secondoftwo
 \fi
}%
\providecommand \natexlab [1]{#1}%
\providecommand \enquote  [1]{``#1''}%
\providecommand \bibnamefont  [1]{#1}%
\providecommand \bibfnamefont [1]{#1}%
\providecommand \citenamefont [1]{#1}%
\providecommand \href@noop [0]{\@secondoftwo}%
\providecommand \href [0]{\begingroup \@sanitize@url \@href}%
\providecommand \@href[1]{\@@startlink{#1}\@@href}%
\providecommand \@@href[1]{\endgroup#1\@@endlink}%
\providecommand \@sanitize@url [0]{\catcode `\\12\catcode `\$12\catcode
  `\&12\catcode `\#12\catcode `\^12\catcode `\_12\catcode `\%12\relax}%
\providecommand \@@startlink[1]{}%
\providecommand \@@endlink[0]{}%
\providecommand \url  [0]{\begingroup\@sanitize@url \@url }%
\providecommand \@url [1]{\endgroup\@href {#1}{\urlprefix }}%
\providecommand \urlprefix  [0]{URL }%
\providecommand \Eprint [0]{\href }%
\providecommand \doibase [0]{https://doi.org/}%
\providecommand \selectlanguage [0]{\@gobble}%
\providecommand \bibinfo  [0]{\@secondoftwo}%
\providecommand \bibfield  [0]{\@secondoftwo}%
\providecommand \translation [1]{[#1]}%
\providecommand \BibitemOpen [0]{}%
\providecommand \bibitemStop [0]{}%
\providecommand \bibitemNoStop [0]{.\EOS\space}%
\providecommand \EOS [0]{\spacefactor3000\relax}%
\providecommand \BibitemShut  [1]{\csname bibitem#1\endcsname}%
\let\auto@bib@innerbib\@empty
\bibitem [{\citenamefont {Onsager}(1949)}]{Onsager1949TheParticles}%
  \BibitemOpen
  \bibfield  {author} {\bibinfo {author} {\bibfnamefont {L.}~\bibnamefont
  {Onsager}},\ }\bibfield  {title} {\bibinfo {title} {{the Effects of Shape on
  the Interaction of Colloidal Particles}},\ }\href
  {https://doi.org/10.1111/j.1749-6632.1949.tb27296.x} {\bibfield  {journal}
  {\bibinfo  {journal} {Annals of the New York Academy of Sciences}\ }\textbf
  {\bibinfo {volume} {51}},\ \bibinfo {pages} {627} (\bibinfo {year}
  {1949})}\BibitemShut {NoStop}%
\bibitem [{\citenamefont {Frenkel}(1991)}]{Frenkel1991}%
  \BibitemOpen
  \bibfield  {author} {\bibinfo {author} {\bibfnamefont {D.}~\bibnamefont
  {Frenkel}},\ }\bibfield  {title} {\bibinfo {title} {{Statistical Mechanics of
  Liquid Crystals}},\ }in\ \href@noop {} {\emph {\bibinfo {booktitle} {Liquids,
  Freezing and the Glass Transition}}},\ \bibinfo {editor} {edited by\ \bibinfo
  {editor} {\bibfnamefont {J.~P.}\ \bibnamefont {Hansen}}, \bibinfo {editor}
  {\bibfnamefont {D.}~\bibnamefont {Levesque}},\ and\ \bibinfo {editor}
  {\bibfnamefont {J.}~\bibnamefont {Zinn-Justin}}}\ (\bibinfo  {publisher}
  {North-Holland, Amsterdam},\ \bibinfo {year} {1991})\ pp.\ \bibinfo {pages}
  {689--762}\BibitemShut {NoStop}%
\bibitem [{\citenamefont {Frenkel}\ \emph {et~al.}(1984)\citenamefont
  {Frenkel}, \citenamefont {Mulder},\ and\ \citenamefont
  {McTague}}]{Frenkel1984}%
  \BibitemOpen
  \bibfield  {author} {\bibinfo {author} {\bibfnamefont {D.}~\bibnamefont
  {Frenkel}}, \bibinfo {author} {\bibfnamefont {B.~M.}\ \bibnamefont
  {Mulder}},\ and\ \bibinfo {author} {\bibfnamefont {J.~P.}\ \bibnamefont
  {McTague}},\ }\bibfield  {title} {\bibinfo {title} {{Phase Diagram of a
  System of Hard Ellipsoids}},\ }\href
  {https://doi.org/10.1103/PhysRevLett.52.287} {\bibfield  {journal} {\bibinfo
  {journal} {Physical Review Letters}\ }\textbf {\bibinfo {volume} {52}},\
  \bibinfo {pages} {287} (\bibinfo {year} {1984})}\BibitemShut {NoStop}%
\bibitem [{\citenamefont {Stroobants}\ \emph {et~al.}(1986)\citenamefont
  {Stroobants}, \citenamefont {Lekkerkerker},\ and\ \citenamefont
  {Frenkel}}]{Stroobants1986}%
  \BibitemOpen
  \bibfield  {author} {\bibinfo {author} {\bibfnamefont {A.}~\bibnamefont
  {Stroobants}}, \bibinfo {author} {\bibfnamefont {H.~N.}\ \bibnamefont
  {Lekkerkerker}},\ and\ \bibinfo {author} {\bibfnamefont {D.}~\bibnamefont
  {Frenkel}},\ }\bibfield  {title} {\bibinfo {title} {{Evidence for Smectic
  Order in a Fluid of Hard Parallel Spherocylinders}},\ }\href
  {https://doi.org/10.1103/PhysRevLett.57.1452} {\bibfield  {journal} {\bibinfo
   {journal} {Physical Review Letters}\ }\textbf {\bibinfo {volume} {57}},\
  \bibinfo {pages} {1452} (\bibinfo {year} {1986})}\BibitemShut {NoStop}%
\bibitem [{\citenamefont {Lipkin}\ and\ \citenamefont
  {Oxtoby}(1983)}]{Lipkin1983ASmectics}%
  \BibitemOpen
  \bibfield  {author} {\bibinfo {author} {\bibfnamefont {M.~D.}\ \bibnamefont
  {Lipkin}}\ and\ \bibinfo {author} {\bibfnamefont {D.~W.}\ \bibnamefont
  {Oxtoby}},\ }\bibfield  {title} {\bibinfo {title} {{A systematic density
  functional approach to the mean field theory of smectics}},\ }\href
  {https://doi.org/10.1063/1.445973} {\bibfield  {journal} {\bibinfo  {journal}
  {The Journal of Chemical Physics}\ }\textbf {\bibinfo {volume} {79}},\
  \bibinfo {pages} {1939} (\bibinfo {year} {1983})}\BibitemShut {NoStop}%
\bibitem [{\citenamefont {Evans}(1992)}]{Evans2007LiquidGeometry}%
  \BibitemOpen
  \bibfield  {author} {\bibinfo {author} {\bibfnamefont {G.~T.}\ \bibnamefont
  {Evans}},\ }\bibfield  {title} {\bibinfo {title} {{Liquid crystal smectic-A
  phases and issues of geometry}},\ }\href
  {https://doi.org/10.1080/00268979200102141} {\bibfield  {journal} {\bibinfo
  {journal} {Molecular Physics}\ }\textbf {\bibinfo {volume} {76}},\ \bibinfo
  {pages} {1359 } (\bibinfo {year} {1992})}\BibitemShut {NoStop}%
\bibitem [{\citenamefont {Wittmann}\ \emph {et~al.}(2014)\citenamefont
  {Wittmann}, \citenamefont {Marechal},\ and\ \citenamefont
  {Mecke}}]{Wittmann2014}%
  \BibitemOpen
  \bibfield  {author} {\bibinfo {author} {\bibfnamefont {R.}~\bibnamefont
  {Wittmann}}, \bibinfo {author} {\bibfnamefont {M.}~\bibnamefont {Marechal}},\
  and\ \bibinfo {author} {\bibfnamefont {K.}~\bibnamefont {Mecke}},\ }\bibfield
   {title} {\bibinfo {title} {{Fundamental measure theory for smectic phases:
  Scaling behavior and higher order terms}},\ }\bibfield  {journal} {\bibinfo
  {journal} {Journal of Chemical Physics}\ }\textbf {\bibinfo {volume} {141}},\
  \href {https://doi.org/10.1063/1.4891326} {10.1063/1.4891326} (\bibinfo
  {year} {2014})\BibitemShut {NoStop}%
\bibitem [{\citenamefont {Wittmann}\ \emph {et~al.}(2017)\citenamefont
  {Wittmann}, \citenamefont {Sitta}, \citenamefont {Smallenburg},\ and\
  \citenamefont {L{\"{o}}wen}}]{Wittmann2017}%
  \BibitemOpen
  \bibfield  {author} {\bibinfo {author} {\bibfnamefont {R.}~\bibnamefont
  {Wittmann}}, \bibinfo {author} {\bibfnamefont {C.~E.}\ \bibnamefont {Sitta}},
  \bibinfo {author} {\bibfnamefont {F.}~\bibnamefont {Smallenburg}},\ and\
  \bibinfo {author} {\bibfnamefont {H.}~\bibnamefont {L{\"{o}}wen}},\
  }\bibfield  {title} {\bibinfo {title} {{Phase diagram of two-dimensional hard
  rods from fundamental mixed measure density functional theory}},\ }\bibfield
  {journal} {\bibinfo  {journal} {Journal of Chemical Physics}\ }\textbf
  {\bibinfo {volume} {147}},\ \href {https://doi.org/10.1063/1.4996131}
  {10.1063/1.4996131} (\bibinfo {year} {2017}),\ \Eprint
  {https://arxiv.org/abs/1708.01248} {arXiv:1708.01248} \BibitemShut {NoStop}%
\bibitem [{\citenamefont {Hosino}\ \emph {et~al.}(1979)\citenamefont {Hosino},
  \citenamefont {Nakano},\ and\ \citenamefont {Kimura}}]{Hosino1979}%
  \BibitemOpen
  \bibfield  {author} {\bibinfo {author} {\bibfnamefont {M.}~\bibnamefont
  {Hosino}}, \bibinfo {author} {\bibfnamefont {H.}~\bibnamefont {Nakano}},\
  and\ \bibinfo {author} {\bibfnamefont {H.}~\bibnamefont {Kimura}},\
  }\bibfield  {title} {\bibinfo {title} {{Nematic-Smectic Transition in an
  Aligned Rod System}},\ }\href {https://doi.org/10.1143/JPSJ.46.1709}
  {\bibfield  {journal} {\bibinfo  {journal} {Journal of the Physical Society
  of Japan}\ }\textbf {\bibinfo {volume} {46}},\ \bibinfo {pages} {1709}
  (\bibinfo {year} {1979})}\BibitemShut {NoStop}%
\bibitem [{\citenamefont {Mulder}(1987)}]{Mulder1987Density-functionalFluid}%
  \BibitemOpen
  \bibfield  {author} {\bibinfo {author} {\bibfnamefont {B.}~\bibnamefont
  {Mulder}},\ }\bibfield  {title} {\bibinfo {title} {{Density-functional
  approach to smectic order in an aligned hard-rod fluid}},\ }\href
  {https://doi.org/10.1103/PhysRevA.35.3095} {\bibfield  {journal} {\bibinfo
  {journal} {Physical Review A}\ }\textbf {\bibinfo {volume} {35}},\ \bibinfo
  {pages} {3095} (\bibinfo {year} {1987})}\BibitemShut {NoStop}%
\bibitem [{\citenamefont {Taylor}\ \emph {et~al.}(1989)\citenamefont {Taylor},
  \citenamefont {Hentschke},\ and\ \citenamefont
  {Herzfeld}}]{Taylor1989TheorySpherocylinders}%
  \BibitemOpen
  \bibfield  {author} {\bibinfo {author} {\bibfnamefont {M.~P.}\ \bibnamefont
  {Taylor}}, \bibinfo {author} {\bibfnamefont {R.}~\bibnamefont {Hentschke}},\
  and\ \bibinfo {author} {\bibfnamefont {J.}~\bibnamefont {Herzfeld}},\
  }\bibfield  {title} {\bibinfo {title} {{Theory of ordered phases in a system
  of parallel hard spherocylinders}},\ }\href
  {https://doi.org/10.1103/PhysRevLett.62.800} {\bibfield  {journal} {\bibinfo
  {journal} {Physical Review Letters}\ }\textbf {\bibinfo {volume} {62}},\
  \bibinfo {pages} {800} (\bibinfo {year} {1989})}\BibitemShut {NoStop}%
\bibitem [{\citenamefont {Gray}\ and\ \citenamefont {Mosley}(1976)}]{Gray1976}%
  \BibitemOpen
  \bibfield  {author} {\bibinfo {author} {\bibfnamefont {G.~W.}\ \bibnamefont
  {Gray}}\ and\ \bibinfo {author} {\bibfnamefont {A.}~\bibnamefont {Mosley}},\
  }\bibfield  {title} {\bibinfo {title} {{Trends in the nematic–isotropic
  liquid transition temperatures for the homologous series of 4-n-alkoxy- and
  4-n-alkyl-4-cyanobiphenyls}},\ }\href {https://doi.org/10.1039/P29760000097}
  {\bibfield  {journal} {\bibinfo  {journal} {J. Chem. Soc., Perkin Trans. 2}\
  }\textbf {\bibinfo {volume} {2}},\ \bibinfo {pages} {97} (\bibinfo {year}
  {1976})}\BibitemShut {NoStop}%
\bibitem [{\citenamefont {Cacelli}\ \emph {et~al.}(2007)\citenamefont
  {Cacelli}, \citenamefont {{De Gaetani}}, \citenamefont {Prampolini},\ and\
  \citenamefont {Tani}}]{Cacelli2007}%
  \BibitemOpen
  \bibfield  {author} {\bibinfo {author} {\bibfnamefont {I.}~\bibnamefont
  {Cacelli}}, \bibinfo {author} {\bibfnamefont {L.}~\bibnamefont {{De
  Gaetani}}}, \bibinfo {author} {\bibfnamefont {G.}~\bibnamefont
  {Prampolini}},\ and\ \bibinfo {author} {\bibfnamefont {A.}~\bibnamefont
  {Tani}},\ }\bibfield  {title} {\bibinfo {title} {{Liquid Crystal Properties
  of the n -Alkyl-cyanobiphenyl Series from Atomistic Simulations with Ab
  Initio Derived Force Fields}},\ }\href {https://doi.org/10.1021/jp065806l}
  {\bibfield  {journal} {\bibinfo  {journal} {The Journal of Physical Chemistry
  B}\ }\textbf {\bibinfo {volume} {111}},\ \bibinfo {pages} {2130} (\bibinfo
  {year} {2007})}\BibitemShut {NoStop}%
\bibitem [{\citenamefont {Belli}\ \emph {et~al.}(2011)\citenamefont {Belli},
  \citenamefont {Patti}, \citenamefont {Dijkstra},\ and\ \citenamefont {van
  Roij}}]{deplete}%
  \BibitemOpen
  \bibfield  {author} {\bibinfo {author} {\bibfnamefont {S.}~\bibnamefont
  {Belli}}, \bibinfo {author} {\bibfnamefont {A.}~\bibnamefont {Patti}},
  \bibinfo {author} {\bibfnamefont {M.}~\bibnamefont {Dijkstra}},\ and\
  \bibinfo {author} {\bibfnamefont {R.}~\bibnamefont {van Roij}},\ }\bibfield
  {title} {\bibinfo {title} {Polydispersity stabilizes biaxial nematic liquid
  crystals},\ }\href {https://doi.org/10.1103/PhysRevLett.107.148303}
  {\bibfield  {journal} {\bibinfo  {journal} {Phys. Rev. Lett.}\ }\textbf
  {\bibinfo {volume} {107}},\ \bibinfo {pages} {148303} (\bibinfo {year}
  {2011})}\BibitemShut {NoStop}%
\bibitem [{\citenamefont {Kamien}(2014)}]{Kamien2014}%
  \BibitemOpen
  \bibfield  {author} {\bibinfo {author} {\bibfnamefont {R.~D.}\ \bibnamefont
  {Kamien}},\ }\bibfield  {title} {\bibinfo {title} {{Entropic Attraction and
  Ordering}},\ }in\ \href {https://doi.org/10.1002/9783527682300.ch1} {\emph
  {\bibinfo {booktitle} {Soft Matter}}},\ \bibinfo {editor} {edited by\
  \bibinfo {editor} {\bibfnamefont {G.}~\bibnamefont {Gompper}}\ and\ \bibinfo
  {editor} {\bibfnamefont {M.}~\bibnamefont {Schick}}}\ (\bibinfo  {publisher}
  {Wiley-VCH Verlag GmbH \& Co. KGaA},\ \bibinfo {address} {Weinheim,
  Germany},\ \bibinfo {year} {2014})\ pp.\ \bibinfo {pages} {1--40}\BibitemShut
  {NoStop}%
\bibitem [{\citenamefont {K\"ohler}(1929)}]{maluma}%
  \BibitemOpen
  \bibfield  {author} {\bibinfo {author} {\bibfnamefont {W.}~\bibnamefont
  {K\"ohler}},\ }\href@noop {} {\emph {\bibinfo {title} {Gestalt Psychology}}}\
  (\bibinfo  {publisher} {H. Liveright},\ \bibinfo {address} {New York},\
  \bibinfo {year} {1929})\BibitemShut {NoStop}%
\bibitem [{\citenamefont {Ramachandran}\ and\ \citenamefont
  {Hubbard}(2001)}]{Ramachandran2001}%
  \BibitemOpen
  \bibfield  {author} {\bibinfo {author} {\bibfnamefont {V.~S.}\ \bibnamefont
  {Ramachandran}}\ and\ \bibinfo {author} {\bibfnamefont {E.~M.}\ \bibnamefont
  {Hubbard}},\ }\bibfield  {title} {\bibinfo {title} {{Synaesthesia - a window
  into perception, thought and language}},\ }\href@noop {} {\bibfield
  {journal} {\bibinfo  {journal} {Journal of Conciousness Studies}\ }\textbf
  {\bibinfo {volume} {8}},\ \bibinfo {pages} {3} (\bibinfo {year}
  {2001})}\BibitemShut {NoStop}%
\bibitem [{\citenamefont {{\'{C}}wiek}\ \emph {et~al.}(2022)\citenamefont
  {{\'{C}}wiek}, \citenamefont {Fuchs}, \citenamefont {Draxler}, \citenamefont
  {Asu}, \citenamefont {Dediu}, \citenamefont {Hiovain}, \citenamefont
  {Kawahara}, \citenamefont {Koutalidis}, \citenamefont {Krifka}, \citenamefont
  {Lippus}, \citenamefont {Lupyan}, \citenamefont {Oh}, \citenamefont {Paul},
  \citenamefont {Petrone}, \citenamefont {Ridouane}, \citenamefont {Reiter},
  \citenamefont {Sch{\"{u}}mchen}, \citenamefont {Szalontai}, \citenamefont
  {{\"{U}}nal-Logacev}, \citenamefont {Zeller}, \citenamefont {Perlman},\ and\
  \citenamefont {Winter}}]{Cwiek2022}%
  \BibitemOpen
  \bibfield  {author} {\bibinfo {author} {\bibfnamefont {A.}~\bibnamefont
  {{\'{C}}wiek}}, \bibinfo {author} {\bibfnamefont {S.}~\bibnamefont {Fuchs}},
  \bibinfo {author} {\bibfnamefont {C.}~\bibnamefont {Draxler}}, \bibinfo
  {author} {\bibfnamefont {E.~L.}\ \bibnamefont {Asu}}, \bibinfo {author}
  {\bibfnamefont {D.}~\bibnamefont {Dediu}}, \bibinfo {author} {\bibfnamefont
  {K.}~\bibnamefont {Hiovain}}, \bibinfo {author} {\bibfnamefont
  {S.}~\bibnamefont {Kawahara}}, \bibinfo {author} {\bibfnamefont
  {S.}~\bibnamefont {Koutalidis}}, \bibinfo {author} {\bibfnamefont
  {M.}~\bibnamefont {Krifka}}, \bibinfo {author} {\bibfnamefont
  {P.}~\bibnamefont {Lippus}}, \bibinfo {author} {\bibfnamefont
  {G.}~\bibnamefont {Lupyan}}, \bibinfo {author} {\bibfnamefont {G.~E.}\
  \bibnamefont {Oh}}, \bibinfo {author} {\bibfnamefont {J.}~\bibnamefont
  {Paul}}, \bibinfo {author} {\bibfnamefont {C.}~\bibnamefont {Petrone}},
  \bibinfo {author} {\bibfnamefont {R.}~\bibnamefont {Ridouane}}, \bibinfo
  {author} {\bibfnamefont {S.}~\bibnamefont {Reiter}}, \bibinfo {author}
  {\bibfnamefont {N.}~\bibnamefont {Sch{\"{u}}mchen}}, \bibinfo {author}
  {\bibfnamefont {{\'{A}}.}~\bibnamefont {Szalontai}}, \bibinfo {author}
  {\bibfnamefont {{\"{O}}.}~\bibnamefont {{\"{U}}nal-Logacev}}, \bibinfo
  {author} {\bibfnamefont {J.}~\bibnamefont {Zeller}}, \bibinfo {author}
  {\bibfnamefont {M.}~\bibnamefont {Perlman}},\ and\ \bibinfo {author}
  {\bibfnamefont {B.}~\bibnamefont {Winter}},\ }\bibfield  {title} {\bibinfo
  {title} {{The bouba/kiki effect is robust across cultures and writing
  systems}},\ }\bibfield  {journal} {\bibinfo  {journal} {Philosophical
  Transactions of the Royal Society B: Biological Sciences}\ }\textbf {\bibinfo
  {volume} {377}},\ \href {https://doi.org/10.1098/rstb.2020.0390}
  {10.1098/rstb.2020.0390} (\bibinfo {year} {2022})\BibitemShut {NoStop}%
\bibitem [{\citenamefont {Doi}\ and\ \citenamefont
  {Edwards}(1986)}]{Doi1986TheDynamics}%
  \BibitemOpen
  \bibfield  {author} {\bibinfo {author} {\bibfnamefont {M.}~\bibnamefont
  {Doi}}\ and\ \bibinfo {author} {\bibfnamefont {S.~F.}\ \bibnamefont
  {Edwards}},\ }\href
  {https://books.google.co.uk/books/about/The_Theory_of_Polymer_Dynamics.html?id=dMzGyWs3GKcC}
  {\emph {\bibinfo {title} {{The Theory of Polymer Dynamics}}}}\ (\bibinfo
  {publisher} {Oxford University Press},\ \bibinfo {year} {1986})\BibitemShut
  {NoStop}%
\bibitem [{\citenamefont {Kirkwood}(1950)}]{Kirkwood1950}%
  \BibitemOpen
  \bibfield  {author} {\bibinfo {author} {\bibfnamefont {J.~G.}\ \bibnamefont
  {Kirkwood}},\ }\bibfield  {title} {\bibinfo {title} {{Critique of the free
  volume theory of the liquid state}},\ }\href
  {https://doi.org/10.1063/1.1747635} {\bibfield  {journal} {\bibinfo
  {journal} {The Journal of Chemical Physics}\ }\textbf {\bibinfo {volume}
  {18}},\ \bibinfo {pages} {380} (\bibinfo {year} {1950})}\BibitemShut
  {NoStop}%
\bibitem [{\citenamefont
  {Edwards}(1967{\natexlab{a}})}]{Edwards1967StatisticalI}%
  \BibitemOpen
  \bibfield  {author} {\bibinfo {author} {\bibfnamefont {S.~F.}\ \bibnamefont
  {Edwards}},\ }\bibfield  {title} {\bibinfo {title} {{Statistical mechanics
  with topological constraints: I}},\ }\href
  {https://doi.org/10.1088/0370-1328/91/3/301} {\bibfield  {journal} {\bibinfo
  {journal} {Proceedings of the Physical Society}\ }\textbf {\bibinfo {volume}
  {91}},\ \bibinfo {pages} {513} (\bibinfo {year}
  {1967}{\natexlab{a}})}\BibitemShut {NoStop}%
\bibitem [{\citenamefont
  {Edwards}(1967{\natexlab{b}})}]{Edwards1967StatisticalII}%
  \BibitemOpen
  \bibfield  {author} {\bibinfo {author} {\bibfnamefont {S.~F.}\ \bibnamefont
  {Edwards}},\ }\bibfield  {title} {\bibinfo {title} {{Statistical mechanics
  with topological constraints: II}},\ }\href
  {https://doi.org/10.1088/0370-1328/91/3/301} {\bibfield  {journal} {\bibinfo
  {journal} {Proceedings of the Physical Society}\ }\textbf {\bibinfo {volume}
  {91}},\ \bibinfo {pages} {513} (\bibinfo {year}
  {1967}{\natexlab{b}})}\BibitemShut {NoStop}%
\bibitem [{\citenamefont {Zwanzig}(1963)}]{zwanzig}%
  \BibitemOpen
  \bibfield  {author} {\bibinfo {author} {\bibfnamefont {R.}~\bibnamefont
  {Zwanzig}},\ }\bibfield  {title} {\bibinfo {title} {First‐order phase
  transition in a gas of long thin rods},\ }\href
  {https://doi.org/10.1063/1.1734518} {\bibfield  {journal} {\bibinfo
  {journal} {The Journal of Chemical Physics}\ }\textbf {\bibinfo {volume}
  {39}},\ \bibinfo {pages} {1714} (\bibinfo {year} {1963})},\ \Eprint
  {https://arxiv.org/abs/https://doi.org/10.1063/1.1734518}
  {https://doi.org/10.1063/1.1734518} \BibitemShut {NoStop}%
\bibitem [{\citenamefont {Tonks}(1936)}]{Tonks1936TheSpheres}%
  \BibitemOpen
  \bibfield  {author} {\bibinfo {author} {\bibfnamefont {L.}~\bibnamefont
  {Tonks}},\ }\bibfield  {title} {\bibinfo {title} {{The complete equation of
  state of one, two and Three-dimensional gases of hard elastic spheres}},\
  }\href {https://doi.org/10.1103/PhysRev.50.955} {\bibfield  {journal}
  {\bibinfo  {journal} {Physical Review}\ }\textbf {\bibinfo {volume} {50}},\
  \bibinfo {pages} {955} (\bibinfo {year} {1936})}\BibitemShut {NoStop}%
\bibitem [{\citenamefont {Ree}\ and\ \citenamefont {Taikyue}(1972)}]{Ree1972}%
  \BibitemOpen
  \bibfield  {author} {\bibinfo {author} {\bibfnamefont {F.~H.}\ \bibnamefont
  {Ree}}\ and\ \bibinfo {author} {\bibfnamefont {R.~E.}\ \bibnamefont
  {Taikyue}},\ }\bibfield  {title} {\bibinfo {title} {{Statistical mechanics of
  the parallel hard squares in canonical ensemble}},\ }\href
  {https://doi.org/10.1063/1.1677059} {\bibfield  {journal} {\bibinfo
  {journal} {The Journal of Chemical Physics}\ }\textbf {\bibinfo {volume}
  {56}},\ \bibinfo {pages} {5434} (\bibinfo {year} {1972})}\BibitemShut
  {NoStop}%
\bibitem [{\citenamefont {Wojciechowski}\ and\ \citenamefont
  {Frenkel}(2004)}]{Wojciechowski2004}%
  \BibitemOpen
  \bibfield  {author} {\bibinfo {author} {\bibfnamefont {K.}~\bibnamefont
  {Wojciechowski}}\ and\ \bibinfo {author} {\bibfnamefont {D.}~\bibnamefont
  {Frenkel}},\ }\bibfield  {title} {\bibinfo {title} {{Tetratic phase in the
  planar hard square system?}},\ }\href
  {https://doi.org/10.12921/cmst.2004.10.02.235-255} {\bibfield  {journal}
  {\bibinfo  {journal} {Computational Methods in Science and Technology}\
  }\textbf {\bibinfo {volume} {10}},\ \bibinfo {pages} {235} (\bibinfo {year}
  {2004})}\BibitemShut {NoStop}%
\bibitem [{\citenamefont {Belli}\ \emph {et~al.}(2012)\citenamefont {Belli},
  \citenamefont {Dijkstra},\ and\ \citenamefont {{Van Roij}}}]{Belli2012}%
  \BibitemOpen
  \bibfield  {author} {\bibinfo {author} {\bibfnamefont {S.}~\bibnamefont
  {Belli}}, \bibinfo {author} {\bibfnamefont {M.}~\bibnamefont {Dijkstra}},\
  and\ \bibinfo {author} {\bibfnamefont {R.}~\bibnamefont {{Van Roij}}},\
  }\bibfield  {title} {\bibinfo {title} {{Free minimization of the fundamental
  measure theory functional: Freezing of parallel hard squares and cubes}},\
  }\bibfield  {journal} {\bibinfo  {journal} {Journal of Chemical Physics}\
  }\textbf {\bibinfo {volume} {137}},\ \href
  {https://doi.org/10.1063/1.4754836} {10.1063/1.4754836} (\bibinfo {year}
  {2012})\BibitemShut {NoStop}%
\bibitem [{\citenamefont {Edwards}\ and\ \citenamefont
  {Freed}(1969)}]{Edwards1969TheI}%
  \BibitemOpen
  \bibfield  {author} {\bibinfo {author} {\bibfnamefont {S.~F.}\ \bibnamefont
  {Edwards}}\ and\ \bibinfo {author} {\bibfnamefont {K.~F.}\ \bibnamefont
  {Freed}},\ }\bibfield  {title} {\bibinfo {title} {{The entropy of a confined
  polymer. I}},\ }\href {https://doi.org/10.1088/0305-4470/2/2/002} {\bibfield
  {journal} {\bibinfo  {journal} {Journal of Physics A: General Physics}\
  }\textbf {\bibinfo {volume} {2}},\ \bibinfo {pages} {145} (\bibinfo {year}
  {1969})}\BibitemShut {NoStop}%
\bibitem [{\citenamefont {Alexander}(1977)}]{Alexander1977}%
  \BibitemOpen
  \bibfield  {author} {\bibinfo {author} {\bibfnamefont {S.}~\bibnamefont
  {Alexander}},\ }\bibfield  {title} {\bibinfo {title} {{Polymer adsorption on
  small spheres. A scaling approach}},\ }\href
  {https://doi.org/10.1051/jphys:01977003808097700} {\bibfield  {journal}
  {\bibinfo  {journal} {Journal de Physique}\ }\textbf {\bibinfo {volume}
  {38}},\ \bibinfo {pages} {977} (\bibinfo {year} {1977})}\BibitemShut
  {NoStop}%
\bibitem [{\citenamefont {{De Gennes}}(1976)}]{DeGennes1976}%
  \BibitemOpen
  \bibfield  {author} {\bibinfo {author} {\bibfnamefont {P.}~\bibnamefont {{De
  Gennes}}},\ }\bibfield  {title} {\bibinfo {title} {{Scaling theory of polymer
  adsorption}},\ }\href {https://doi.org/10.1051/jphys:0197600370120144500}
  {\bibfield  {journal} {\bibinfo  {journal} {Journal de Physique}\ }\textbf
  {\bibinfo {volume} {37}},\ \bibinfo {pages} {1445} (\bibinfo {year}
  {1976})}\BibitemShut {NoStop}%
\bibitem [{\citenamefont {de~Gennes}(1980)}]{DeGennes1980}%
  \BibitemOpen
  \bibfield  {author} {\bibinfo {author} {\bibfnamefont {P.~G.}\ \bibnamefont
  {de~Gennes}},\ }\bibfield  {title} {\bibinfo {title} {{Conformations of
  Polymers Attached to an Interface}},\ }\href
  {https://doi.org/10.1021/ma60077a009} {\bibfield  {journal} {\bibinfo
  {journal} {Macromolecules}\ }\textbf {\bibinfo {volume} {13}},\ \bibinfo
  {pages} {1069} (\bibinfo {year} {1980})}\BibitemShut {NoStop}%
\bibitem [{\citenamefont {Milner}\ \emph {et~al.}(1988)\citenamefont {Milner},
  \citenamefont {Witten},\ and\ \citenamefont {Cates}}]{Milner1988TheoryBrush}%
  \BibitemOpen
  \bibfield  {author} {\bibinfo {author} {\bibfnamefont {S.~T.}\ \bibnamefont
  {Milner}}, \bibinfo {author} {\bibfnamefont {T.~A.}\ \bibnamefont {Witten}},\
  and\ \bibinfo {author} {\bibfnamefont {M.~E.}\ \bibnamefont {Cates}},\
  }\bibfield  {title} {\bibinfo {title} {{Theory of the Grafted Polymer
  Brush}},\ }\href {https://doi.org/10.1021/ma00186a051} {\bibfield  {journal}
  {\bibinfo  {journal} {Macromolecules}\ }\textbf {\bibinfo {volume} {21}},\
  \bibinfo {pages} {2610} (\bibinfo {year} {1988})}\BibitemShut {NoStop}%
\bibitem [{\citenamefont {Abramowitz}\ and\ \citenamefont
  {Stegun}(1964)}]{Abramowitz1964HandbookTables}%
  \BibitemOpen
  \bibfield  {author} {\bibinfo {author} {\bibfnamefont {M.}~\bibnamefont
  {Abramowitz}}\ and\ \bibinfo {author} {\bibfnamefont {I.~A.}\ \bibnamefont
  {Stegun}},\ }\href@noop {} {\emph {\bibinfo {title} {{Handbook of
  Mathematical Functions with Formulas, Graphs, and Mathematical Tables}}}}\
  (\bibinfo  {publisher} {Dover Publications},\ \bibinfo {address} {New York
  City},\ \bibinfo {year} {1964})\BibitemShut {NoStop}%
\end{thebibliography}

\end{document}